\documentclass[twocolumn,showpacs,preprintnumbers,superscriptaddress,amsmath,amssymb,PRL]{revtex4}
\usepackage{graphicx}
\usepackage{dcolumn}
\usepackage{float}
\usepackage{mathptmx, courier, pifont}
\usepackage[scaled=0.92]{helvet}
\usepackage[T1]{fontenc}
\usepackage{textcomp}
\usepackage{color}
\usepackage{booktabs}

\usepackage[dvipsnames]{xcolor}
\usepackage[colorlinks=true,urlcolor=Blue,linkcolor=Blue]{hyperref}
\usepackage[all]{hypcap}

\begin{document}

\title{Masses of ground and isomeric states of $^{101}$In and configuration-dependent shell evolution in odd-$A$ indium isotopes}

\author{X.~Xu}
\affiliation{Key Laboratory of High Precision Nuclear Spectroscopy and Center for Nuclear Matter Science, Institute of Modern Physics, Chinese Academy of Sciences, Lanzhou 730000, China}
\affiliation{School of Science, Xi$^{\prime}$an Jiaotong University, Xi$^{\prime}$an 710049, China}
\author{J.~H.~Liu}
\affiliation{Key Laboratory of High Precision Nuclear Spectroscopy and Center for Nuclear Matter Science, Institute of Modern Physics, Chinese Academy of Sciences, Lanzhou 730000,  China}
\affiliation{School of Nuclear Science and Technology, University of Chinese Academy of Sciences, Beijing 100049, China}
\author{C.~X.~Yuan}
\affiliation{Sino-French Institute of Nuclear Engineering and Technology, Sun Yat-Sen University, Zhuhai 519082,  China}
\author{Y.~M.~Xing}
\affiliation{Key Laboratory of High Precision Nuclear Spectroscopy and Center for Nuclear Matter Science, Institute of Modern Physics, Chinese Academy of Sciences, Lanzhou 730000, China}
\author{M.~Wang}\thanks{Corresponding author:  wangm@impcas.ac.cn} 
\affiliation{Key Laboratory of High Precision Nuclear Spectroscopy and Center for Nuclear Matter Science, Institute of Modern Physics, Chinese Academy of Sciences, Lanzhou 730000, China}
\affiliation{School of Nuclear Science and Technology, University of Chinese Academy of Sciences, Beijing 100049, China}
\author{Y.~H.~Zhang}\thanks{Corresponding author: yhzhang@impcas.ac.cn}
\affiliation{Key Laboratory of High Precision Nuclear Spectroscopy and Center for Nuclear Matter Science, Institute of Modern Physics, Chinese Academy of Sciences, Lanzhou 730000,  China}
\affiliation{School of Nuclear Science and Technology, University of Chinese Academy of Sciences, Beijing 100049, China}
\author{X.~H.~Zhou} 
\affiliation{Key Laboratory of High Precision Nuclear Spectroscopy and Center for Nuclear Matter Science, Institute of Modern Physics, Chinese Academy of Sciences, Lanzhou 730000,  China}
\affiliation{School of Nuclear Science and Technology, University of Chinese Academy of Sciences, Beijing 100049, China}
\author{Yu.~A.~Litvinov}
\affiliation{Key Laboratory of High Precision Nuclear Spectroscopy and Center for Nuclear Matter Science, Institute of Modern Physics, Chinese Academy of Sciences, Lanzhou 730000,  China}
\affiliation{GSI Helmholtzzentrum f{\"u}r Schwerionenforschung, Planckstra{\ss}e 1, Darmstadt, 64291 Germany}
\author{K.~Blaum}
\affiliation{Max-Planck-Institut f\"{u}r Kernphysik, Saupfercheckweg 1, 69117 Heidelberg, Germany}
\author{R.~J.~Chen}
\affiliation{Key Laboratory of High Precision Nuclear Spectroscopy and Center for Nuclear Matter Science, Institute of Modern Physics, Chinese Academy of Sciences, Lanzhou 730000,  China}
\author{X.~C.~Chen}
\affiliation{Key Laboratory of High Precision Nuclear Spectroscopy and Center for Nuclear Matter Science, Institute of Modern Physics, Chinese Academy of Sciences, Lanzhou 730000,  China}
\author{C.~Y.~Fu}
\affiliation{Key Laboratory of High Precision Nuclear Spectroscopy and Center for Nuclear Matter Science, Institute of Modern Physics, Chinese Academy of Sciences, Lanzhou 730000,  China}
\author{B.~S.~Gao}
\affiliation{Key Laboratory of High Precision Nuclear Spectroscopy and Center for Nuclear Matter Science, Institute of Modern Physics, Chinese Academy of Sciences, Lanzhou 730000,  China}
\affiliation{School of Nuclear Science and Technology, University of Chinese Academy of Sciences, Beijing 100049, China}
\author{J.~J.~He}
\affiliation{Key Laboratory of Beam Technology of Ministry of Education, College of Nuclear Science and Technology, Beijing Normal University, Beijing 100875, China}
\affiliation{Key Laboratory of High Precision Nuclear Spectroscopy and Center for Nuclear Matter Science, Institute of Modern Physics, Chinese Academy of Sciences, Lanzhou 730000,  China}
\author{S.~Kubono}
\affiliation{Key Laboratory of High Precision Nuclear Spectroscopy and Center for Nuclear Matter Science, Institute of Modern Physics, Chinese Academy of Sciences, Lanzhou 730000,  China}
\author{Y.~H.~Lam}
\affiliation{Key Laboratory of High Precision Nuclear Spectroscopy and Center for Nuclear Matter Science, Institute of Modern Physics, Chinese Academy of Sciences, Lanzhou 730000,  China}
\author{H.~F.~Li}
\affiliation{Key Laboratory of High Precision Nuclear Spectroscopy and Center for Nuclear Matter Science, Institute of Modern Physics, Chinese Academy of Sciences, Lanzhou 730000,  China}
\affiliation{School of Nuclear Science and Technology, University of Chinese Academy of Sciences, Beijing 100049, China}
\author{M.~L.~Liu}
\affiliation{Key Laboratory of High Precision Nuclear Spectroscopy and Center for Nuclear Matter Science, Institute of Modern Physics, Chinese Academy of Sciences, Lanzhou 730000,  China}
\affiliation{School of Nuclear Science and Technology, University of Chinese Academy of Sciences, Beijing 100049, China}
\author{X.~W.~Ma}
\affiliation{Key Laboratory of High Precision Nuclear Spectroscopy and Center for Nuclear Matter Science, Institute of Modern Physics, Chinese Academy of Sciences, Lanzhou 730000,  China}
\affiliation{School of Nuclear Science and Technology, University of Chinese Academy of Sciences, Beijing 100049, China}
\author{P.~Shuai}
\affiliation{Key Laboratory of High Precision Nuclear Spectroscopy and Center for Nuclear Matter Science, Institute of Modern Physics, Chinese Academy of Sciences, Lanzhou 730000,  China}
\author{M.~Si}
\affiliation{Key Laboratory of High Precision Nuclear Spectroscopy and Center for Nuclear Matter Science, Institute of Modern Physics, Chinese Academy of Sciences, Lanzhou 730000,  China}
\affiliation{School of Nuclear Science and Technology, University of Chinese Academy of Sciences, Beijing 100049, China}
\author{M.~Z.~Sun}
\affiliation{Key Laboratory of High Precision Nuclear Spectroscopy and Center for Nuclear Matter Science, Institute of Modern Physics, Chinese Academy of Sciences, Lanzhou 730000,  China}
\affiliation{School of Nuclear Science and Technology, University of Chinese Academy of Sciences, Beijing 100049, China}
\author{X.~L.~Tu}
\affiliation{Key Laboratory of High Precision Nuclear Spectroscopy and Center for Nuclear Matter Science, Institute of Modern Physics, Chinese Academy of Sciences, Lanzhou 730000,  China}
\affiliation{School of Nuclear Science and Technology, University of Chinese Academy of Sciences, Beijing 100049, China}
\author{Q.~Wang}
\affiliation{Key Laboratory of High Precision Nuclear Spectroscopy and Center for Nuclear Matter Science, Institute of Modern Physics, Chinese Academy of Sciences, Lanzhou 730000,  China}
\affiliation{School of Nuclear Science and Technology, University of Chinese Academy of Sciences, Beijing 100049, China}
\author{H.~S.~Xu}
\affiliation{Key Laboratory of High Precision Nuclear Spectroscopy and Center for Nuclear Matter Science, Institute of Modern Physics, Chinese Academy of Sciences, Lanzhou 730000,  China}
\affiliation{School of Nuclear Science and Technology, University of Chinese Academy of Sciences, Beijing 100049, China}
\author{X.~L.~Yan}
\affiliation{Key Laboratory of High Precision Nuclear Spectroscopy and Center for Nuclear Matter Science, Institute of Modern Physics, Chinese Academy of Sciences, Lanzhou 730000,  China}
\author{J.~C.~Yang}
\affiliation{Key Laboratory of High Precision Nuclear Spectroscopy and Center for Nuclear Matter Science, Institute of Modern Physics, Chinese Academy of Sciences, Lanzhou 730000,  China}
\affiliation{School of Nuclear Science and Technology, University of Chinese Academy of Sciences, Beijing 100049, China}
\author{Y.~J.~Yuan}
\affiliation{Key Laboratory of High Precision Nuclear Spectroscopy and Center for Nuclear Matter Science, Institute of Modern Physics, Chinese Academy of Sciences, Lanzhou 730000,  China}
\affiliation{School of Nuclear Science and Technology, University of Chinese Academy of Sciences, Beijing 100049, China}
\author{Q.~Zeng}
\affiliation{Key Laboratory of High Precision Nuclear Spectroscopy and Center for Nuclear Matter Science, Institute of Modern Physics, Chinese Academy of Sciences, Lanzhou 730000,  China}
\affiliation{School of Nuclear Science and Engineering, East China University of Technology, Nanchang, 330013, China}
\author{P.~Zhang}
\affiliation{Key Laboratory of High Precision Nuclear Spectroscopy and Center for Nuclear Matter Science, Institute of Modern Physics, Chinese Academy of Sciences, Lanzhou 730000,  China}
\affiliation{School of Nuclear Science and Technology, University of Chinese Academy of Sciences, Beijing 100049, China}
\author{X.~Zhou}
\affiliation{Key Laboratory of High Precision Nuclear Spectroscopy and Center for Nuclear Matter Science, Institute of Modern Physics, Chinese Academy of Sciences, Lanzhou 730000,  China}
\affiliation{School of Nuclear Science and Technology, University of Chinese Academy of Sciences, Beijing 100049, China}
\author{W.~L.~Zhan}
\affiliation{Key Laboratory of High Precision Nuclear Spectroscopy and Center for Nuclear Matter Science, Institute of Modern Physics, Chinese Academy of Sciences, Lanzhou 730000,  China}
\author{S.~Litvinov}
\affiliation{GSI Helmholtzzentrum f{\"u}r Schwerionenforschung, Planckstra{\ss}e 1, Darmstadt, 64291 Germany}
\author{G.~Audi}
\affiliation{CSNSM, Univ Paris-Sud, CNRS/IN2P3, Universit\'{e} Paris-Saclay, 91405 Orsay, France}
\author{S.~Naimi}
\affiliation{RIKEN Nishina Center, RIKEN, Saitama 351-0198, Japan}
\author{T.~Uesaka}
\affiliation{RIKEN Nishina Center, RIKEN, Saitama 351-0198, Japan}
\author{Y.~Yamaguchi}
\affiliation{RIKEN Nishina Center, RIKEN, Saitama 351-0198, Japan}
\author{T.~Yamaguchi}
\affiliation{Department of Physics, Saitama University, Saitama 338-8570, Japan}
\author{A.~Ozawa}
\affiliation{Insititute of Physics, University of Tsukuba, Ibaraki 305-8571, Japan}
\author{B.~H.~Sun}
\affiliation{School of Physics and Nuclear Energy Engineering, Beihang University, Beijing 100191,  China}
\author{K.~Kaneko}
\affiliation{Department of Physics, Kyushu Sangyo University, Fukuoka 813-8503, Japan}
\author{Y.~Sun}
\affiliation{School of Physics and Astronomy, Shanghai Jiao Tong University, Shanghai 200240,  China}
\affiliation{Key Laboratory of High Precision Nuclear Spectroscopy and Center for Nuclear Matter Science, Institute of Modern Physics, Chinese Academy of Sciences, Lanzhou 730000, China}
\author{F.~R.~Xu}
\affiliation{State Key Laboratory of Nuclear Physics and Technology, School of Physics, Peking University, Beijing 100871,  China}

\date{\today}

\begin{abstract}

We report first precision mass measurements of the $1/2^-$ isomeric and $9/2^+$ ground states of $^{101}$In. The determined isomeric excitation energy continues a smooth trend of odd-$A$ indium isotopes up to the immediate vicinity of $N=50$ magic number. This trend can be confirmed by dedicated shell model calculations only if the neutron configuration mixing is considered. We find that the single particle energies are different for different states of the same isotope. The presented configuration-dependent shell evolution, type II shell evolution, in odd-$A$ nuclei is discussed for the first time. Our results will facilitate future studies of single-particle neutron states.


\pacs{21.10.Dr, 21.10.Pc, 27.60.+j}
\end{abstract}
\maketitle

Properties of the nuclides around the doubly-magic nucleus $^{100}$Sn have attracted intense research efforts both in experiment and theory \cite{100SnPPNP}. 
Of particular interest are the isomeric states which provide unique insight into the nuclear structure. 
Many isomers were predicted and some of them have been observed around $^{100}$Sn \cite{2,3,4,5,6,7,8,9,10}. 
Various implications from the obtained results on nuclear structure \cite{2,3,4,5,6,7,8,9} and on astrophysics \cite{10} have been extensively discussed. 
Currently, all known isomers in this region were discovered by using decay spectroscopy.

In odd-$A$ indium isotopes ($Z$ = 49), the ground states have \emph{J}$^{\pi}$ = 9/2$^{+}$ with the $\pi (1g_{9/2})^{-1}$ proton-hole character. 
When promoting a proton from $\pi 2p_{1/2}$ orbital to $\pi 1g_{9/2}$, the $\pi (2p_{1/2})^{-1}$ proton-hole state is formed. 
Due to the slow $M$4 transition from the 1/2$^{-}$ state to the 9/2$^{+}$ ground state, the former state is always a $\beta$-decaying isomer. On the one hand, the excitation energies of such isomers are directly related to the energy gap between the $\pi 2p_{1/2}$ and $\pi 1g_{9/2}$ orbitals, and thus allow the study of the gap evolution as a function of neutron number. On the other hand, proton occupations inside the same nucleus in either $\pi 2p_{1/2}$ or $\pi 1g_{9/2}$ single-particle orbital may affect, via the proton-neutron ($p$-$n$) interactions, the ordering of neutron orbitals, e.~g., $\nu 1g_{7/2}$ and $\nu 2d_{5/2}$. 
It has been shown~\cite{Yusu2014,Kre2016} that the monopole part of the $p$-$n$ interaction plays a crucial role for the shell evolution within the same nucleus, the so-called type II shell evolution. Such a newly reported shell evolution can
be investigated in odd-$A$ indium isotopes 
since indium has just one proton less than the magic tin ($Z=50$) and has thus relatively pure proton particle-hole excitations. 

When approaching $^{100}$Sn, $^{103}$In is the most neutron-deficient indium isotope in which the 1/2$^{-}$ isomeric state, $^{\rm 103m}$In, was found \cite{8}. 
The level structure of $^{101}$In was studied via the $^{50}$Cr($^{58}$Ni, 3p4n)$^{101}$In fusion-evaporation reaction \cite{8}
and the $\beta$-decay of $^{101}$Sn \cite{101Snbeta}, but $^{\rm 101m}$In was not observed in these experiments. 
It was estimated in Ref.~\cite{8} that the dominant decay mode of the yet unknown $^{\rm 101m}$In is beta decay. 
There has been a circumstantial evidence of the existence of the 1/2$^{-}$ isomer in $^{97}$In, though its excitation energy and half-life could not be determined~\cite{97In}.

The production of the 1/2$^{-}$ state may be less favored in nuclear reactions due to the low spin. Therefore more sensitive methods should be developed for the identification of the isomer.
To investigate nuclear isomers, the state-of-the-art mass spectrometry~\cite{Klaus06} gained importance in recent years thanks to the improved mass resolution, which is high enough to directly resolve low-lying isomers from the corresponding nuclear ground states. 
New isomers have been discovered with Penning trap \cite{12} as well as storage-ring mass spectrometry \cite{13a,13,13b,14}. 
In such experiments, the excitation energy is measured directly as the mass difference between the isomer and the ground state. 
For already known isomers, mass spectrometry can independently provide information on the isomer excitation energies \cite{15,16,17}. 


In this paper, we report the direct identification of the 1/2$^{-}$ isomer in $^{101}$In. 
Masses of the isomeric and ground states, and thus automatically the isomer excitation energy, $E_x(1/2^-){}$, of $^{101}$In are determined for the first time.

The experiment was conducted at the Cooler Storage Ring (CSR) accelerator complex of the Heavy Ion Research Facility in Lanzhou (HIRFL) located at the Institute of Modern Physics, P. R. China.
It has been done in a similar way as our previous experiments \cite{16,TuPRL,TuNIMA,ZhangPRL,ZhangPRC,XingPLB}.
A 400.88 MeV/u $^{112}$Sn$^{35+}$ primary beam was accelerated and accumulated in the main storage ring CSRm, operating as a heavy-ion synchrotron. 
Every 25 seconds, the beam was fast extracted and focused onto a $\sim$10-mm-thick beryllium target placed at the entrance of the in-flight separator RIBLL2~\cite{Xia jiawen}. 
The projectile fragments were selected and analyzed by RIBLL2. A cocktail beam of $10-30$ ions per spill was injected into the experimental storage ring CSRe. 
The CSRe was tuned into the isochronous ion-optical mode with the transition energy of  ${\gamma}_{t}=1.302$. 
In this mode the revolution times of the stored ions depend in first order only on their mass-to-charge ratios, which is the basis of the Isochronous Mass Spectrometry (IMS) \cite{MHa,FGM}.
Both RIBLL2 and CSRe were set to a fixed magnetic rigidity of \emph{B}${\rho}=5.3374$ Tm to optimize the transmission of the \emph{T}$_{z}=3/2$ nuclides centered on $^{101}$In$^{49+}$. 
Other nuclides within the acceptance of the RIBLL2-CSRe system of about $\pm0.2$\% were also transmitted and stored in CSRe. 
In order to achieve a better mass resolving power, a 50 mm wide slit was introduced~\cite{slit2019} in the dispersive straight section of CSRe to reduce the momentum spread of the secondary beam.

\begin{figure}[htbp]	
	\center
	\includegraphics[width=0.9\columnwidth]{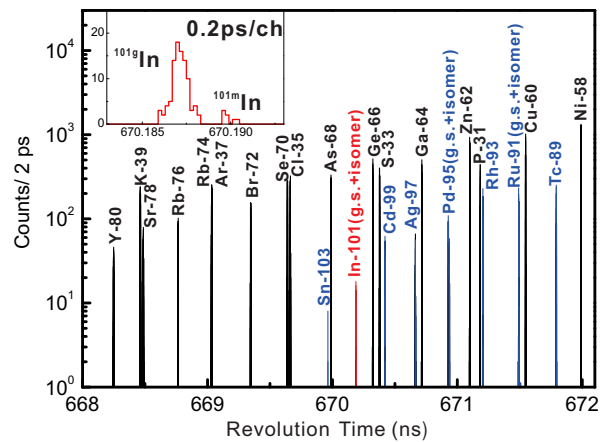}
	\caption{(color online) Part of the revolution time spectrum zoomed in a time window of 668 ns $\leq$ \emph{t}  $\leq$ 672 ns. The nuclides with well-known masses (black color) were used as calibrants in the mass calibration.
	The insert shows the well-resolved peaks of the ground and isomeric states of $^{101}$In.
	}
	\label{spectrum}
\end{figure}

To measure the revolution times of stored ions, a high-performance time-of-flight detector (ToF) 
equipped with a 19 $\mu$g/cm$^{2}$ carbon foil of 40 mm in diameter was installed inside CSRe \cite{MeiNIMA}. 
The time resolution of the detector was about 50 ps. 
Secondary electrons were released from the foil each time when an ion passed through the carbon foil. 
The electrons were guided to a microchannel plate (MCP) by a perpendicularly arranged electric and magnetic field. 
The timing signals from the MCP detector were directly sampled with a digital oscilloscope at a sampling rate of 40~GHz. 
For each injection of ions into CSRe the recording time was set to 200 $\mu$s corresponding to $\sim300$ revolutions of the ions. Those ions which circulated in CSRe for more than about 75 revolutions (50 $\mu$s) were considered in the data analysis. 
The revolution time of each individual ion was extracted from the measured periodic timing signals \cite{TuNIMA}. 
The entire data analyses, including the time shift correction and particle identification, have been conducted following the procedures described in Refs.~\cite{TuNIMA,ZhangPRC,XingNIMA}. Some details of the present analysis were reported in Ref.~\cite{NPR}. The final revolution time spectrum has been obtained by accumulating all events.

Figure~\ref{spectrum} presents a part of the revolution time spectrum zoomed in at a time window of 668 ns $\leq$ \emph{t}  $\leq$ 672 ns. 
In this time range, the minimum standard deviation, $\sigma_T$, of the peaks was about 0.5 ps 
corresponding to a mass resolving power of \emph{m}/${\Delta}$\emph{m}${\sim}$ $3.2 \times 10^5$ (FWHM). 
The insert in Fig.~\ref{spectrum} illustrates the revolution time spectrum expanded around $^{101}$In. 
In addition to the main peak corresponding to the ground state of $^{101}$In, 
seven counts are observed at a mean revolution time of $\sim$5$\sigma_T$ larger than the time of the main peak ($\sim $2.8 ps).
These counts are attributed to the particles other than the ground state of $^{101}$In. 
Given the fact, that all particles in Fig.~\ref{spectrum} have been unambiguously identified 
as belonging to different series with certain $T_z=({Z-N})/{2}$, and their yields and revolution times for the isotopes of same $T_z$ 
follow the expected systematics as a function of mass number, the above-mentioned seven counts can be assigned only to an unknown isomer in $^{101}$In.

\begin{table*}[htbp]
	\caption{
	Mass excess $(ME)$ values  of the ground ($^{101}$In) and isomeric ($^{101m}$In) states obtained in the present work. 
	Also given are the numbers of counts, the widths of the revolution time peaks ($\sigma_T$), spin parities ($J^{\pi}$), 
	the differences between the experimental and extrapolated values ($\Delta ME$), the measured $(E_x^{\rm{CSRe}}$($1/2^-$)$)$ and extrapolated $(E_x^{\rm{AME16}}$($1/2^-$)$)$ excitation energies. 
	The corresponding extrapolated values from AME$^{\prime}$16 \cite{AME16} are marked with label '\#'. 
	 }
\begin{tabular*}{0.95\textwidth}{lcccccccc}
	\toprule
		~~Atom~~     & ~~Counts~~   &  ~~$\sigma_T$ ~~ & $J^{\pi}$     &~~ $ME_{\rm{CSRe}}$  ~~& ~~$ME_{\rm{AME16}}$ ~~    & ~~~$\Delta ME$  ~~~   &        ~~~ $E_x^{\rm{CSRe}}(1/2^-)$ ~~~  &         ~~~$E_x^{\rm{AME16}}(1/2^-)$ ~~~   \vspace{1mm}  \\
            &     &  (ps)    & $(\hbar)$   & (keV)  &  (keV)  & (keV)   &   (keV)   &   (keV)         \\	
	
		\hline   \vspace{-2mm} \\
~~$^{101}$In            & $       95       $    &  ~$0.54$~     &~ $9/2^+$\#~     & ~$-68550(13)(6)$~  &  $-68610(200)$\#   & $60(200)$   &           0          &           0             \\
~~$^{\rm 101m}$In          & $      7         $    &   ~$0.54$~     & ~$1/2^-$\#~     & ~$-67891(48)(6)$~  & $-68060(220)$\#     & $169(225)$   &    $ 659(50)$  &  $550(100)$\# \\
		\bottomrule
	\end{tabular*}
	\label{tablemass}
\end{table*}

Most of the nuclides in Fig.~\ref{spectrum} have well-known masses. 
Their mass excess $(ME)$ values from the latest Atomic Mass Evaluation, AME$^{\prime}$16,~\cite{AME16} 
were used to fit their mass-to-charge ratios $m/q$ versus their corresponding revolution times $T$.
A third order polynomial function has been employed. 
The mass calibration has been checked by re-determining the $ME$ values of each of the $N_c$ reference nuclides ($N_c$ = $17$) by using the other $N_c~-~1$ ones as calibrants.
The normalized $\chi_{n}$ defined as:
\begin{eqnarray}\label{Chi-square equ}	
\chi_{n}=\sqrt{\frac{1}{N_c}\sum\limits_{i=1}^{N_c}\frac{[(\frac{m}{q})_{i,\rm{exp}} - (\frac{m}{q})_{i,\rm{AME}}]^{2}}{\sigma^2_{{i,\rm{exp}}} + \sigma^2_{{i,\rm{AME}}}}}~,
\end{eqnarray}
was found to be $\chi_{n}$ = $1.363$. 
This value is slightly outside the expected range of $\chi_{n}$ = $1\pm0.171$ at $1\sigma$ confidence level.
A systematic error of 6 keV has been added to the corresponding statistical uncertainties. 
The mass excess values of $^{101}$In and its isomer determined in this work are presented in Table \ref{tablemass}.

\begin{figure}[htbp]	
  	\flushleft
  	\includegraphics[width=1.0\columnwidth]{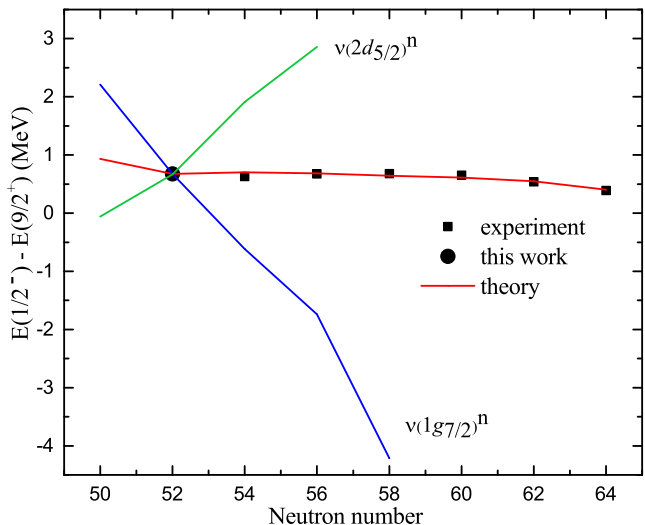}
  	\caption{(color online) Excitation energies of the $1/2^-$ isomers in odd-$A$ indium isotopes together with shell model calculations. 
	The filled squares are experiment data from ENSDF \cite{NNDC} and the filled circle for the present work. 
	The green line is obtained by adding valence neutrons only to the $\nu 2d_{5/2}$ orbital, while the blue line corresponds to the $\nu 1g_{7/2}$ occupation only. 
	The red line is calculated within the model space of $\nu 1g_{7/2}, \nu 2d_{5/2}, \nu 2d_{3/2}, \nu 3s_{1/2}$, and $\nu 1h_{11/2}$ orbitals. }
  	\label{101In}
  \end{figure}

Based on the systematics, a low-lying $1/2^-$ isomer $^{\rm 101m}$In was predicted to have a half-life comparable to the ground state~\cite{AME16}. 
The excitation energy determined in the present work, $659(50)$ keV, is in a reasonable agreement with the 
extrapolation of AME$^{\prime}$16~\cite{AME16} and the shell model calculations~\cite{101Snbeta}. 
Following the systematics of $1/2^-$ isomers in odd-$A$ indium isotopes, see Fig.~\ref{101In}, 
a spin-and-parity of $1/2^-$ is assigned here to the new isomer in $^{101}$In. 

The available experimental data on the excitation energies of the $1/2^-$ states, $E_x(1/2^-)$, in odd-$A$ indium isotopes extend now to the immediate vicinity of the doubly-magic nucleus $^{100}$Sn. These energies provide a direct information on the relative position of the $\pi 1g_{9/2}$ and $\pi2p_{1/2}$ single particle orbitals, i.~e. the $Z=40$ sub-shell gap. 
It has been shown~\cite{Ag2019} that the size of the $Z=40$ subshell gap is strongly affected by the spin-isospin part of the proton-neutron interaction.
To get more insight into the structure of the $9/2^+$ ground state and the $1/2^-$ isomer in
$^{101}$In as well as the microscopic origin of the $Z=40$ sub-shell evolution in this region, 
dedicated shell-model calculations have been performed by using the KSHELL code \cite{kshell} 
with the state-of-the-art monopole-based universal interaction $V_{\rm MU}$ plus a spin-orbit force from M3Y($V_{\rm MU}+LS$) \cite{Tensor2, M3Y}. 
The $V_{\rm MU}$ interaction consists of a Gaussian central force and a tensor force \cite{Tensor2} and has been successfully applied to describe the
shell structure of exotic nuclei in various regions \cite{application1,application2,application3,application4,application5}. 
In our calculations, 
the model space for protons consisted of the $\pi 1f_{5/2}, \pi 2p_{3/2}, \pi 2p_{1/2}$, and $\pi 1g_{9/2}$ orbitals. 
For neutrons, three model spaces were considered, namely the $\nu 2d_{5/2}$ single orbital, the $\nu 1g_{7/2}$ single orbital, 
and the $\nu 2d_{5/2}, \nu 1g_{7/2}, \nu 3s_{1/2}, \nu 1h_{11/2}$, and $\nu 2d_{3/2}$ five orbitals. 
In all calculations, $^{78}$Ni was taken as an inert core and 
the single particle energies were tuned in order to give a consistent $E_x(1/2^-)$ value for $^{101}$In. 
The calculated results are presented in Fig.~\ref{101In}.
We see from this figure that if the neutrons are restricted only to the $\nu 1g_{7/2}$ orbital, 
the $E(1/2^-)-E(9/2^+)$ energy difference rapidly decreases from $^{101}$In to $^{107}$In (the blue line in Fig.~\ref{101In}). 
An inverse trend is observed if the neutrons are added to the $\nu 2d_{5/2}$ orbital (the green line in Fig.~\ref{101In}).
This indicates that the calculations with pure neutron configurations can not reproduce 
the systematics of the experimental energy differences between the $9/2^+$ and $1/2^-$ states.

\begin{figure}[htbp]	
	\flushleft
	\includegraphics[width=1.0\columnwidth]{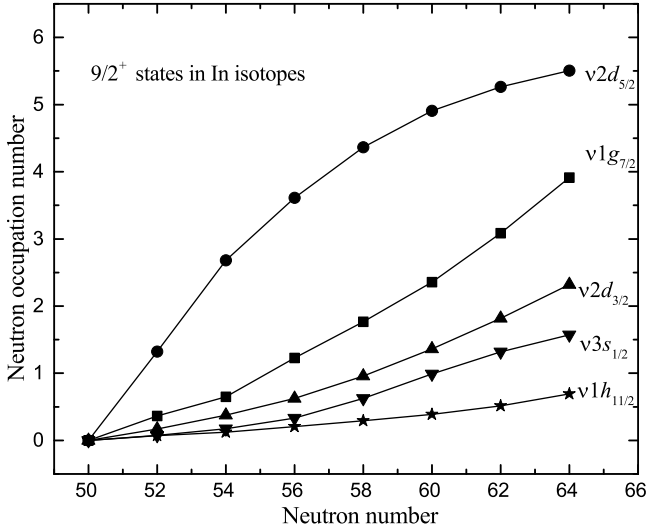}
	\caption{ Neutron occupation numbers for the $9/2^+$ ground states in odd-$A$ indium isotopes from $^{101}$In to $^{113}$In.}
	\label{occupy92}
\end{figure}

\begin{figure}[t]	
	\flushleft
	\includegraphics[width=1.0\columnwidth]{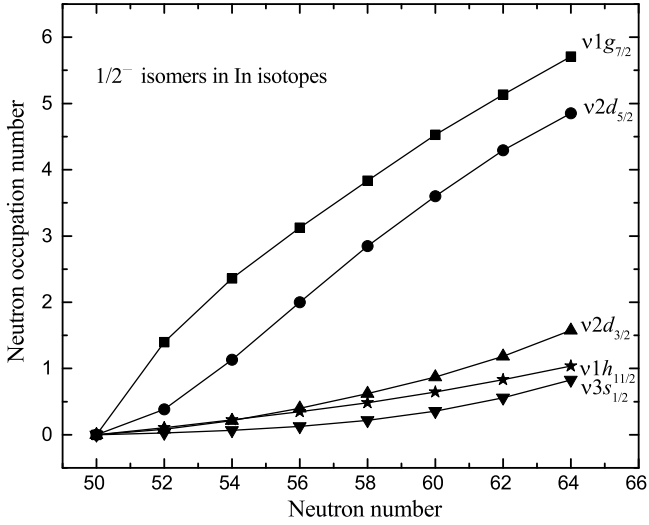}
	\caption{ Neutron occupation numbers for the $1/2^-$ states in odd-$A$ indium isotopes from $^{101}$In to $^{113}$In.}
	\label{occupy12}
\end{figure}

When the model space containing five neutron orbitals is considered in the calculations, 
the experimental trends of energy spacings can be well reproduced by the theory as shown by the red line in Fig.~\ref{101In}. 
On the one hand, this result indicates that the new $1/2^-$ isomer in $^{101}$In is formed in a similar proton-hole configuration as the other isomers in odd-$A$ indium isotopes. On the other hand, the neutron configuration mixing should be considered in understanding the microscopic structure of these states.

 	Based on the properties of the spin-isospin part of the proton-neutron interaction discussed in Ref.~\cite{otsuka2001,Tensor1,Tensor2}, 
	it is known that the interactions between a $ j=l+1/2$ proton with a $j^{\prime}=l^{\prime}-1/2$ neutron is more attractive than the one for the $ j=l+1/2$ proton with a $j^{\prime}=l^{\prime}+1/2$ neutron and vice versa. 
 From this scenario, one may expect that the relative position of $\nu 1g_{7/2}$ and $\nu 2d_{5/2}$ orbitals in each indium isotope depends on the proton configuration. 
 The $1/2^-$ isomer and the $9/2^{+}$ ground state correspond to $\pi (2p_{1/2})^{-1}$ and  $\pi (1g_{9/2})^{-1}$ configurations, respectively. With respect to the $9/2^+$ ground state, the accumulated effects for the $1/2^-$ isomer, due to one proton gain in $\pi 1g_{9/2}$ and one proton loss in $\pi 2p_{1/2}$, can lead to the reduction of energy spacing or even an inversion of the $\nu 1g_{7/2}$ and $\nu 2d_{5/2}$ orbitals.
 

\begin{figure}[htbp]	
	\flushleft
	\includegraphics[width=1.0\columnwidth]{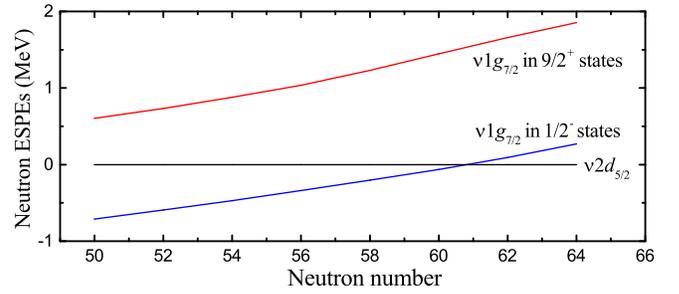}
	\caption{(color online) Neutron effective single particle energies of $\nu 1g_{7/2}$ with respect to $\nu 2d_{5/2}$ for $9/2^+$ and $1/2^-$ states in indium isotopes as calculated with the $V_{MU} + LS$ interaction.}
	\label{espe}
\end{figure}

Figures~\ref{occupy92} and~\ref{occupy12} show the neutron occupation numbers extracted from our calculations 
with the neutron model space of $\nu 2d_{5/2}, \nu 1g_{7/2}, \nu 3s_{1/2}, \nu 1h_{11/2}$, and $\nu 2d_{3/2}$.
For the $9/2^+$ ($1/2^-$) state throughout the region of interest, the valence neutrons occupy mainly the $\nu 2d_{5/2}$ ($\nu 1g_{7/2}$) orbital 
with considerable contributions from the $\nu 1g_{7/2}$ ($\nu 2d_{5/2}$) neutrons. 
The neutron configuration mixing in the $\nu 2d_{5/2}$ and $\nu 1g_{7/2}$ orbitals gives a flat trend of $E_x(1/2^-)$ in $^{101}$In through $^{113}$In. 
Otsuka $et~al.$ have shown~\cite{otsuka2001,Tensor1,Tensor2} that, by adding protons to the $\pi 1g_{9/2}$ orbital, 
the $\nu 2d_{5/2}$ and $\nu 1g_{7/2}$ orbitals become linearly closer, and that they are almost degenerate when approaching $Z=50$. 
The close-lying $\nu 2d_{5/2}$ and $\nu 1g_{7/2}$ orbitals favor neutron configuration mixing as results from our calculations. 

Furthermore, as can be seen in Figs.~\ref{occupy92} and~\ref{occupy12} the neutron occupation of the $\nu 2d_{5/2}$ orbital 
is larger than that of the $\nu 1g_{7/2}$ orbital for the $9/2^+$ ground state, while this ordering is inverted for the $1/2^-$ isomers. As discussed before, the different neutron occupancies for the $9/2^+$ and $1/2^-$ states in each indium isotope indicate that the $\nu 2d_{5/2}$ and $\nu 1g_{7/2}$ single particle energies may be different in the two states of the same nucleus. From $^{99}$In to $^{113}$In, Fig.~\ref{espe} shows that the $\nu 1g_{7/2}$ neutron effective single particle energies (ESPE) in the $9/2^+$ states are $\sim$1 MeV shifted downwards in the $1/2^-$ states if taking ESPE of $\nu 2d_{5/2}$ as reference. It is worth emphasizing that the ordering of the $\nu 2d_{5/2}$ and the $\nu 1g_{7/2}$ orbitals changes for the $1/2^-$ states at around $^{109}$In, while for the $9/2^+$ ground states the ESPE of $\nu 1g_{7/2}$ is larger than that of $\nu 2d_{5/2}$ in all considered indium isotopes.


As shown in Fig.~\ref{101In}, adding neutrons to the $\nu 1g_{7/2}$ ($\nu 2d_{5/2}$) orbital induces lowering (lifting) 
of the $\pi 1g_{9/2}$ orbital as compared to the $\pi2p_{1/2}$ orbital. 
Similarly, in each indium isotope the $\nu 1g_{7/2}$ ESPE becomes lower with respect to the $\nu 2d_{5/2}$ orbital 
when a proton is moved from $\pi2p_{1/2}$ to $\pi 1g_{9/2}$. 
Recently, type II shell evolution has experimentally been confirmed 
in the neutron-rich even-even nuclide $^{96}$Zr \cite{Kre2016} and the odd-odd nuclide $^{70}$Co \cite{70Co}. 
Here, a new example of type II shell evolution is suggested to take place in neutron-deficient odd-$A$ indium isotopes, 
which is induced by the position of the single proton hole. 
The strong many body correlations among two proton orbitals with $j_1 = l_1 + 1/2$ and $j_2 = l_2 - 1/2$ 
and two neutron orbitals with $j_3 = l_3 + 1/2$ and $j_4 = l_4 - 1/2$ contribute to the present results of both, 
type I and II, shell evolution in neutron-deficient indium isotopes. Further cases may exist in other regions of the nuclear chart.

In summary, by using the storage-ring based isochronous mass spectrometry, 
we have measured for the first time the masses of the $1/2^-$ isomer and the $9/2^+$ ground state of $^{101}$In. This extends the systematics of excitation energies of the $1/2^-$ isomers in indium isotopes approaching $N=50$. 
The similar excitation energies of the $1/2^-$ isomers indicate a stable $Z=40$ subshell gap from $^{101}$In to $^{113}$In.
Our state-of-the-art shell-model calculations with $V_{\rm MU}$ plus a spin-orbit force can well describe the available experimental data, and show that the strong configuration mixing of valence neutrons play a key role in explaining the smooth trend of $E_x(1/2^-)$ as a function of neutron number.  
Furthermore, our calculations show that the energies of the $\nu 1g_{7/2}$ and $\nu 2d_{5/2}$ orbitals are different in the $1/2^-$ and $9/2^+$ states of the same indium isotope. This observation is valid for all considered here indium isotopes.
Such configuration-dependent shell evolution, the so called type II shell evolution, studied in the present work
will prompt further studies of single-neutron states by means of ${\beta}$-decay and transfer reaction experiments using radioactive-isotope beams.  

We thank the staff of the accelerator division of the IMP for providing the stable beam. 
This work was supported in part by 
the National Key R\&D Program of China (Grant No. 2018YFA0404401 and No. 2016YFA0400504), 
the Key Research Program of Frontier Sciences of CAS (Grant No. QYZDJ-SSW-S), 
the NSFC (Grants No. 11605249, 11605248, 11605252, 11605267, 11575112, 11835001, 11711540016 and 11775316), 
the Helmholtz-CAS Joint Research Group HCJRG-108, 
Deutscher Akademischer Austauschdienst (DAAD), Programm des Projektbezogenen Personenaustauschs (PPP) with China (Project ID 57389367),
and
the CAS External Cooperation Program (Grant No. GJHZ1305).
Y.A.L. is indebted to the European Research Council (ERC) for support under the European Union's Horizon 2020 Research and Innovation Programme (Grant Agreement No. 682841 "ASTRUm").
Y.H.Z. acknowledges support by the ExtreMe Matter Institute EMMI at the GSI Helmholtzzentrum f{\"u}r Schwerionenforschung, Darmstadt, Germany. 
X.L.T. acknowledges support from the Max Planck Society through the Max-Planck Partner Group.

\end{document}